\documentclass[12pt]{article}
\usepackage{amssymb}
\usepackage{graphicx}
\begin{document}
\newcommand{\beq}{\begin{equation}}
\newcommand{\eeq}{\end{equation}}
\newcommand{\beqa}{\begin{eqnarray}}
\newcommand{\eeqa}{\end{eqnarray}}
\newcommand{\beqar}{\begin{eqnarray*}}
\newcommand{\eeqar}{\end{eqnarray*}}
\newcommand{\al}{\alpha}
\newcommand{\be}{\beta}
\newcommand{\del}{\delta}
\newcommand{\D}{\Delta}
\newcommand{\eps}{\epsilon}
\newcommand{\ga}{\gamma}
\newcommand{\Ga}{\Gamma}
\newcommand{\ka}{\kappa}
\newcommand{\nn}{\nonumber}
\newcommand{\inn}{\!\cdot\!}
\newcommand{\h}{\eta}
\newcommand{\ii}{\iota}
\newcommand{\kk}{\varphi}
\newcommand\F{{}_3F_2}
\newcommand{\la}{\lambda}
\newcommand{\La}{\Lambda}
\newcommand{\na}{\prt}
\newcommand{\Om}{\Omega}
\newcommand{\om}{\omega}
\newcommand{\p}{\phi}
\newcommand{\sig}{\sigma}
\renewcommand{\t}{\theta}
\newcommand{\z}{\zeta}
\newcommand{\ssc}{\scriptscriptstyle}
\newcommand{\eg}{{\it e.g.,}\ }
\newcommand{\ie}{{\it i.e.,}\ }
\newcommand{\labell}[1]{\label{#1}} 
\newcommand{\reef}[1]{(\ref{#1})}
\newcommand\prt{\partial}
\newcommand\veps{\varepsilon}
\newcommand{\pol}{\varepsilon}
\newcommand\vp{\varphi}
\newcommand\ls{\ell_s}
\newcommand\cF{{\cal F}}
\newcommand\cA{{\cal A}}
\newcommand\cS{{\cal S}}
\newcommand\cT{{\cal T}}
\newcommand\cV{{\cal V}}
\newcommand\cL{{\cal L}}
\newcommand\cM{{\cal M}}
\newcommand\cN{{\cal N}}
\newcommand\cG{{\cal G}}
\newcommand\cH{{\cal H}}
\newcommand\cI{{\cal I}}
\newcommand\cJ{{\cal J}}
\newcommand\cl{{\iota}}
\newcommand\cP{{\cal P}}
\newcommand\cQ{{\cal Q}}
\newcommand\cg{{\it g}}
\newcommand\cR{{\cal R}}
\newcommand\cB{{\cal B}}
\newcommand\cO{{\cal O}}
\newcommand\tcO{{\tilde {{\cal O}}}}
\newcommand\bg{\bar{g}}
\newcommand\bb{\bar{b}}
\newcommand\bH{\bar{H}}
\newcommand\bF{\bar{F}}
\newcommand\bX{\bar{X}}
\newcommand\bK{\bar{K}}
\newcommand\bA{\bar{A}}
\newcommand\bZ{\bar{Z}}
\newcommand\bxi{\bar{\xi}}
\newcommand\bphi{\bar{\phi}}
\newcommand\bpsi{\bar{\psi}}
\newcommand\bprt{\bar{\prt}}
\newcommand\bet{\bar{\eta}}
\newcommand\btau{\bar{\tau}}
\newcommand\hF{\hat{F}}
\newcommand\hA{\hat{A}}
\newcommand\hT{\hat{T}}
\newcommand\htau{\hat{\tau}}
\newcommand\hD{\hat{D}}
\newcommand\hf{\hat{f}}
\newcommand\hg{\hat{g}}
\newcommand\hp{\hat{\phi}}
\newcommand\hi{\hat{i}}
\newcommand\ha{\hat{a}}
\newcommand\hb{\hat{b}}
\newcommand\hQ{\hat{Q}}
\newcommand\hP{\hat{\Phi}}
\newcommand\hS{\hat{S}}
\newcommand\hX{\hat{X}}
\newcommand\tL{\tilde{\cal L}}
\newcommand\hL{\hat{\cal L}}
\newcommand\tG{{\widetilde G}}
\newcommand\tg{{\widetilde g}}
\newcommand\tphi{{\widetilde \phi}}
\newcommand\tPhi{{\widetilde \Phi}}
\newcommand\te{{\tilde e}}
\newcommand\tk{{\tilde k}}
\newcommand\tf{{\tilde f}}
\newcommand\ta{{\tilde a}}
\newcommand\tb{{\tilde b}}
\newcommand\tR{{\tilde R}}
\newcommand\teta{{\tilde \eta}}
\newcommand\tF{{\widetilde F}}
\newcommand\tK{{\widetilde K}}
\newcommand\tE{{\widetilde E}}
\newcommand\tpsi{{\tilde \psi}}
\newcommand\tX{{\widetilde X}}
\newcommand\tD{{\widetilde D}}
\newcommand\tO{{\widetilde O}}
\newcommand\tS{{\tilde S}}
\newcommand\tB{{\widetilde B}}
\newcommand\tA{{\widetilde A}}
\newcommand\tT{{\widetilde T}}
\newcommand\tC{{\widetilde C}}
\newcommand\tV{{\widetilde V}}
\newcommand\thF{{\widetilde {\hat {F}}}}
\newcommand\Tr{{\rm Tr}}
\newcommand\tr{{\rm tr}}
\newcommand\STr{{\rm STr}}
\newcommand\hR{\hat{R}}
\newcommand\MZ{\mathbb{Z}}
\newcommand\MR{\mathbb{R}}
\newcommand\M[2]{M^{#1}{}_{#2}}

\newcommand\bS{\textbf{ S}}
\newcommand\bI{\textbf{ I}}
\newcommand\bJ{\textbf{ J}}

\begin{titlepage}
\begin{center}

\vskip 2 cm
{\LARGE \bf    Odd-Derivative  Couplings in \\  \vskip 0.25 cm Heterotic Theory  
 }\\
\vskip 1.25 cm
 Mohammad R. Garousi\footnote{garousi@um.ac.ir}
 
\vskip 1 cm
{{\it Department of Physics, Faculty of Science, Ferdowsi University of Mashhad\\}{\it P.O. Box 1436, Mashhad, Iran}\\}
\vskip .1 cm
 \end{center}
\begin{abstract}

In this paper, our focus is on exploring the gauge-invariant basis for bosonic couplings within the framework of heterotic string theories, specifically examining 3-, 5-, and 7-derivative terms. We thoroughly analyze the invariance of these couplings under T-duality transformations and make a notable observation: the T-duality constraint enforces the vanishing of these couplings. We speculate that this result likely holds true for all higher odd-derivative couplings as well. This is unlike the result in type I superstring theory, where, for example, the couplings of 5 Yang-Mills field strengths are non-zero. 
The vanishing of couplings is consistent with the $O(d,d+16)$ symmetry of the cosmological reduction of the effective action.

\end{abstract}
\end{titlepage}

\section{Introduction} \label{intro}

The classical effective action of string theory exhibits a global $O(d,d,\mathbb{R})$ symmetry at all orders of derivatives when the effective action is dimensionally reduced on a torus $T^{(d)}$ \cite{Sen:1991zi,Hassan:1991mq,Hohm:2014sxa}. The non-geometric subgroup of this symmetry allows us to construct the covariant and gauge-invariant effective action of string theory at any order of derivatives. The construction involves first determining the independent basis at each order of derivatives with arbitrary coupling constants and then imposing T-duality symmetry to determine the coupling constants in each basis.
For circular reduction, this method has been employed in previous works, such as \cite{Garousi:2019mca, Garousi:2023kxw, Garousi:2020gio}, to derive the Neveu-Schwarz-Neveu-Schwarz (NS-NS) couplings up to eight-derivative orders. More recently, a truncated T-duality transformation has been shown to be applicable in finding the Yang-Mills (YM) couplings, as well as NS-NS couplings, in the heterotic theory \cite{Garousi:2024avb}.

In constructing covariant NS-NS bases, it is crucial to account for all contractions of the NS-NS field strengths at each derivative order. Subsequently, one may eliminate from the list of couplings those that are redundant, meaning they can be removed via field redefinitions, integration by parts, and application of Bianchi identities, to find the independent couplings in the minimal scheme  \cite{Metsaev:1987zx}. The circular reduction of such redundant couplings is also invariant under T-duality \cite{Garousi:2019mca, Garousi:2023kxw, Garousi:2020gio}. This symmetry of the redundant couplings may result from the Buscher transformations for the NS-NS fields being linear in the reduction scheme introduced in \cite{Maharana:1992my}.
However, when T-duality is extended to include YM fields, the corresponding Buscher transformations become nonlinear \cite{Bergshoeff:1995cg}. This may cause the redundant couplings to no longer be invariant under the full T-duality. In such cases, one should consider the couplings in the maximal scheme, where the redundant couplings due to field redefinitions are not removed.
In this paper, we first assume the redundant couplings in the presence of YM fields are invariant under the truncated T-duality transformations  \cite{Garousi:2024avb}. 
Hence, to address the coupling constants, we apply the truncated T-duality transformations to the minimal bases. Furthermore, in the Discussion section, we will explore applying the same truncated T-duality transformations to the maximal bases in order to fix their coupling constants.

To enforce the Bianchi identities for the covariant NS-NS couplings, one can utilize the diffeomorphism symmetry to select a specific gauge where the Levi-Civita connection is zero, while its derivatives are non-zero. This gauge selection greatly simplifies the identification of independent gravity couplings within the given basis \cite{Garousi:2019cdn}.
Similarly, in order to impose the Bianchi identities for the YM gauge-invariant couplings, the YM gauge symmetry can be employed to choose a specific gauge where the YM connection is zero, while its derivatives are non-zero. This gauge choice aids in determining the independent YM couplings within the basis, effectively eliminating the commutator term in the YM field strength \cite{Garousi:2024avb}.

To validate the elimination of commutator terms, even for the derivatives of the YM field strength, we employ the following reasoning. When the YM field is present, the derivative of the field strength involves both the Levi-Civita and YM connections. This derivative is given by $\tilde{\nabla}F = \nabla F + \frac{1}{\sqrt{\alpha'}}[A, F]$, where $\nabla$ represents the ordinary covariant derivative involving only the Levi-Civita connection. 
We assume that the YM field is dimensionless and use the fact that each derivative should come with a factor of $\sqrt{\alpha'}$ to make the corresponding coupling dimensionless. Therefore, in the chosen gauge where $A=0$, the first derivative of the YM field strength simplifies to the ordinary covariant derivative.

The second derivative of the YM field strength can be expressed as the sum of symmetric and antisymmetric derivatives:
\begin{equation}
\tilde{\nabla}\tilde{\nabla} F = [\tilde{\nabla}, \tilde{\nabla}]F + \{\tilde{\nabla}, \tilde{\nabla}\}F\,.
\end{equation}
The symmetric part involves $\{\nabla, \nabla\} F$, along with other terms involving the YM gauge field $A$ without derivatives, which become zero in the chosen gauge. On the other hand, the antisymmetric part satisfies the following Bianchi identity:\begin{equation}
[\tilde{\nabla}, \tilde{\nabla}]F = [\nabla, \nabla]F + \frac{1}{\sqrt{\alpha'}}[F,F]\,.
\end{equation}
This indicates that either the couplings on the left-hand side are independent, or the couplings on the right-hand side are independent. The second term on the right-hand side has fewer derivatives than the other terms in this equation.
On the other hand, in our procedure for finding the basis, we assume a derivative expansion for the bases and first construct the independent bases at the 2-derivative level. We then proceed to construct the bases at 3-derivatives, 4-derivatives, and so on. Consequently, the terms $FF$ on the right-hand side are already chosen as independent couplings in the basis at the lower derivative order. Therefore, these terms can only modify the coefficient of the independent terms that have already been chosen as independent couplings. As a result, the $FF$ terms can be removed from the antisymmetric part, and the second derivatives reduce to just the ordinary covariant derivatives. This pattern continues for all higher derivatives of the YM field strength.

Hence, in finding the basis at any order of derivatives, we use the YM gauge symmetry to utilize the following YM field strength and $B$-field strength:
\beqa
F_{\mu\nu}{}^{ij} &=& \partial_\mu A_{\nu}{}^{ij} - \partial_\nu A_{\mu}{}^{ij}, \nn\\
H_{\mu\nu\rho} &=& 3\partial_{[\mu}B_{\nu\rho]} - \frac{3}{2}A_{[\mu}{}^{ij}F_{\nu\rho]}{}_{ij}, \label{FH}
\eeqa
and their ordinary covariant derivatives. Here, the YM gauge field is defined as $A_{\mu}{}^{ij} = A_{\mu}{}^I(\lambda^I)^{ij}$, where the antisymmetric matrices $(\lambda^I)^{ij}$ represent the adjoint representation of the gauge group $SO(32)$ or $E_8\times E_8$ with the normalization $(\lambda^I)^{ij}(\lambda^J)_{ij}=\delta^{IJ}$. 
Note that whenever $H$ appears in the basis, one can set the second term in the above field strength expression to zero in our chosen gauge. However, the second term has a non-zero contribution for the derivative of $H$, which should be taken into account.

Once the basis at the $n$-derivative order is found, it is important to also include the Lorentz Chern-Simons three-form $\Omega_{\mu\nu\rho}$ in the $B$-field strength. This three-form arises as a result of the Green-Schwarz mechanism \cite{Green:1984sg}. Its inclusion leads to couplings at derivative orders higher than $n$ with the same coefficients as those in the basis with $n$ derivatives.
Moreover, to establish a basis suitable for unfixed YM gauges, one should replace the ordinary covariant derivatives with covariant derivatives that incorporate both the Levi-Civita and YM connections. Additionally, the commutator term in the YM field strength should be replaced as follows:
\begin{equation}
F_{\mu\nu}{}^{ij} = \partial_\mu A_{\nu}{}^{ij} - \partial_\nu A_{\mu}{}^{ij} + \frac{1}{\sqrt{\alpha'}}[A_{\mu}{}^{ik},A_{\nu k}{}^{j}]\,.
\end{equation}
This replacement should also be applied to the covariant derivatives of the YM field strength.

Determining the coupling constants in the bases involves applying the non-geometric T-duality constraint to the reduction of the effective action. If the parent couplings are in unfixed YM gauges, the reduced couplings become covariant and invariant under YM gauge transformations in the base space. We can then leverage these symmetries to choose gauges in which the Levi-Civita connection and YM connections in the base space vanish, while their derivatives do not. This allows us to apply the logic described earlier in the base space, effectively eliminating all commutators in the YM field strength and its derivatives in the bases, resulting in ordinary covariant derivatives.
On the other hand, if the parent couplings are in the chosen gauge where $A=0$, the reduced couplings do not contain commutator terms, and the derivatives in the base space are already in the form of ordinary covariant derivatives. 
In our approach for finding the coupling constants, we follow the latter case where the parent couplings and their circular reductions are in the chosen gauge where the commutator terms are zero, and the derivatives in both spacetime and in the base space are treated as ordinary covariant derivatives. Using the fact that the coupling constants are gauge invariant, the resulting coupling constants should be valid for any other gauges as well.

In the context of circular reduction, the non-geometric aspects of T-duality transformations at the leading order of derivatives are described by the Buscher transformations \cite{Buscher:1987sk,Buscher:1987qj}. These transformations, along with their higher derivative generalizations, play a crucial role in determining the coupling constants. Within the reduction scheme \cite{Maharana:1992my,Bergshoeff:1995cg}, the Buscher transformations exhibit linearity for the NS-NS fields but introduce nonlinearity for the YM field in the base space. Specifically, the scalar component of the YM gauge field appears nonlinearly in the corresponding Buscher transformation \cite{Bergshoeff:1995cg}. Furthermore, this scalar field manifests in a nonlinear manner during the circular reduction of the effective actions.
It is a nontrivial task to find higher-derivative couplings that are invariant under the full nonlinear Buscher transformations and their higher-derivative extensions.

A previous study \cite{Garousi:2024avb} introduced a truncation approach for generalized Buscher transformations and the reduced action. This truncation involves removing terms that include derivatives of the scalar field, as well as terms with second and higher orders of the scalar field. It was proposed that these truncated Buscher transformations and reduced action contain sufficient information to determine the coupling constants at each derivative order. The validity of this proposal was confirmed by evaluating the four-derivative couplings in the heterotic theory, which were found to be consistent with the corresponding couplings obtained using the S-matrix method, as reported in the literature  \cite{Gross:1986mw,Kikuchi:1986rk}. Similar truncation schemes for the NS-NS fields were also observed in \cite{Garousi:2019mca}, where the removal of the Levi-Civita connection and all its derivatives in the base space still retains enough information in the T-duality constraint to determine all NS-NS couplings in a basis.
In that case, it would be straightforward to include the Levi-Civita connection and its derivatives in the base space by writing the resulting couplings in the base space in covariant form. That step would not produce any further constraint on the original couplings. However, it would be a nontrivial task to include the terms in the base space that involve the scalar component of the YM field and its derivatives nonlinearly. This step would presumably produce further constraints on the original couplings, which may fix the scheme of the original couplings. 
In other words, we expect the couplings to be invariant only in a specific scheme under the full nonlinear T-duality transformations.
In this paper, we adopt the truncated T-duality transformation introduced in  \cite{Garousi:2024avb} to investigate the couplings involving odd derivatives in the heterotic theory.

The paper is structured as follows:
In Section 2, we explore bases that incorporate 3-, 5-, and 7-derivatives, encompassing both NS-NS and YM field strengths, along with their covariant derivatives. These bases include arbitrary coupling constants.
In Section 3, we employ truncated T-duality techniques to explicitly determine the coupling constants in each basis. We find that T-duality fixes all the coupling constants in these bases to be zero. The result for the 3-derivative coupling is consistent with the one obtained from the S-matrix method. While higher odd-derivative couplings are extremely difficult to explicitly calculate, we speculate that all couplings involving odd-derivatives, which include field strengths and their derivatives, are zero. In other words, the couplings associated with odd derivatives are not YM gauge invariant and should arise from the commutator terms present in the even-derivative couplings of field strengths.
Section 4 provides a concise discussion of our findings and their implications. In particular, we clarify in this section that the vanishing of the odd-derivative couplings is consistent with the $O(d,d+16)$ symmetry of the cosmological reduction of the heterotic effective action.
Throughout our calculations, we utilize the "xAct" package \cite{Nutma:2013zea} for computational purposes.

\section{Minimal Bases}\label{sec.2}

In this section, we focus on identifying the bases that involve 3-, 5-, and 7-derivative orders, incorporating both the NS-NS and YM field strengths. It becomes evident that the couplings in these bases must necessarily include the YM field strength. Without the presence of the YM field strength, it would not be possible to construct couplings with odd-derivatives solely from the NS-NS field strengths.

At the 3-derivative order, we find a single coupling given by:
\beqa
 {\bf L}^{({1/2})}&=&a_1\sqrt{\alpha'}F_\alpha{}^\gamma{}_i{}^j F^{\alpha\beta i k} F_{\beta\gamma k j}\,=\,a_1\sqrt{\alpha'}\,\Tr(F_\alpha{}^\gamma F_{\beta\gamma}F^{\alpha\beta})\,.\labell{L3}
 \eeqa
Here, $a_1$ represents the coupling constant.

At the 5-derivative order, we encounter a total of 51 couplings. However, by eliminating redundancies arising from field redefinitions, integration by parts, and employing the Bianchi identity, we discover that there are 23 independent couplings remaining. Hence, the minimal basis should have 23 couplings. These couplings in a particular scheme are given by: (See the Appendix for a step-by-step explanation of how to find them.)
\beqa
{\bf L}^{({3/2})}&\!\!\!\!=\!\!\!\!&\alpha'\sqrt{\alpha'}\Big[b_{1}   F_{\alpha  }{}^{\epsilon  im } 
F^{\alpha  \beta  jk} F_{\beta  }{}^{\gamma  }{}_{i}{}^{l} F_{
\gamma  \delta  km } F_{\epsilon  }{}^{\delta  
}{}_{jl} +   
 b_{2}   F_{\alpha  }{}^{\epsilon  }{}_{il} F^{\alpha  
\beta  il} F_{\beta  }{}^{\varepsilon  jk} F_{\epsilon  
}{}^{\gamma  }{}_{j}{}^{m } F_{\varepsilon  \gamma  
km }\nn\\&\!\!\!\!\!\!\!\!& +   
 b_{3}   F_{\alpha  \beta  il} F^{\alpha  \beta  il} 
F_{\epsilon  }{}^{\gamma  }{}_{j}{}^{m } F^{\epsilon  \varepsilon  jk} F_{\varepsilon  \gamma  km } +   
 b_{4}   F_{\alpha  }{}^{\epsilon  }{}_{i}{}^{m } 
F^{\alpha  \beta  ik} F_{\beta  }{}^{\nu  
jl} F_{\epsilon  }{}^{\gamma  }{}_{jl} F_{\nu  \gamma  km } \nn\\&\!\!\!\!\!\!\!\!&+   
 b_{5}   F_{\alpha  }{}^{\epsilon  }{}_{i}{}^{l} 
F^{\alpha  \beta  ij} F_{\beta  }{}^{\nu  
}{}_{j}{}^{k} F_{\epsilon  }{}^{\gamma  }{}_{k}{}^{m } 
F_{\nu  \gamma  lm } +   
 b_{6}   F_{\alpha  }{}^{\epsilon  }{}_{i}{}^{j} 
F^{\alpha  \beta  il} F_{\beta  }{}^{\nu  
}{}_{j}{}^{k} F_{\epsilon  }{}^{\gamma  }{}_{k}{}^{m } 
F_{\nu  \gamma  lm } \nn\\&\!\!\!\!\!\!\!\!&+   
 b_{7}   F_{\alpha  }{}^{\epsilon  }{}_{i}{}^{k} 
F^{\alpha  \beta  ij} F_{\beta  }{}^{\nu  
}{}_{j}{}^{l} F_{\epsilon  }{}^{\gamma  }{}_{k}{}^{m } 
F_{\nu  \gamma  lm } +   
 b_{8}   F_{\alpha  \beta  i}{}^{j} F^{\alpha  \beta  
ik} F_{\varepsilon  }{}^{\gamma  }{}_{j}{}^{m } 
F^{\varepsilon  \nu  }{}_{k}{}^{l} 
F_{\nu  \gamma  lm }\nn\\&\!\!\!\!\!\!\!\!& +   
 b_{9}   F_{\alpha  \beta  }{}^{lk} F^{\alpha  \beta  
ji} F_{\gamma  \delta  km } F_{\nu  }{}^{\delta  }{}_{i}{}^{m } F^{\nu  \gamma  }{}_{jl} +   
 b_{10}   F_{\alpha  \beta  }{}^{ik} F^{\alpha  \beta  
jl} F_{\gamma  \delta  km } F_{\varepsilon\varepsilon  }{}^{\delta  }{}_{i}{}^{m } F^{\nu  \gamma  }{}_{jl} \labell{L5}\\&\!\!\!\!\!\!\!\!&+   
 b_{11}   F_{\alpha  }{}^{\epsilon  }{}_{i}{}^{k} 
F^{\alpha  \beta  ij} F^{\nu  \gamma  
}{}_{jk} H_{\beta  \nu  }{}^{\delta  } 
H_{\epsilon  \gamma  \delta  } +   
 b_{12}   F^{\alpha  \beta  ij} F^{\gamma  \delta  
}{}_{jk} F^{\epsilon  \varepsilon  }{}_{i}{}^{k} H_{\alpha  
\beta  \epsilon  } H_{\varepsilon  \gamma  \delta  } \nn\\&\!\!\!\!\!\!\!\!&+   
 b_{13}   F_{\alpha  }{}^{\epsilon  }{}_{i}{}^{k} 
F^{\alpha  \beta  ij} F^{\nu  \gamma  
}{}_{jk} H_{\beta  \epsilon  }{}^{\delta  } H_{\nu  \gamma  \delta  } +   
 b_{14}   F_{\alpha  }{}^{\epsilon  }{}_{i}{}^{k} 
F^{\alpha  \beta  ij} F_{\beta  }{}^{\nu  
}{}_{jk} H_{\epsilon  }{}^{\gamma  \delta  } H_{\nu  \gamma  \delta  }\nn\\&\!\!\!\!\!\!\!\!& +   
 b_{15}   F_{\alpha  }{}^{\epsilon  }{}_{i}{}^{k} 
F^{\alpha  \beta  ij} F_{\beta  \epsilon  jk} H_{\nu  \gamma  \delta  } H^{\nu  
\gamma  \delta  } +   
 b_{16}   F_{\alpha  }{}^{\epsilon  }{}_{i}{}^{k} 
F^{\alpha  \beta  ij} F^{\nu  \gamma  
}{}_{jk} R_{\beta  \epsilon  \nu  
\gamma  } \nn\\&\!\!\!\!\!\!\!\!&+   
 b_{17}   F_{\alpha  }{}^{\beta  ji} F_{\epsilon  
}{}^{\gamma  }{}_{ik} F^{\epsilon  \varepsilon  }{}_{j}{}^{k} H_{
\beta  \varepsilon  \gamma  } \nabla^{\alpha  }\Phi  +   
 b_{18}   F_{\alpha  }{}^{\beta  ij} F_{\beta  
}{}^{\epsilon  }{}_{i}{}^{k} F^{\nu  \gamma  
}{}_{jk} H_{\epsilon  \nu  \gamma  } 
\nabla^{\alpha  }\Phi  \nn\\&\!\!\!\!\!\!\!\!&+   
 b_{19}   F_{\beta  }{}^{\varepsilon  }{}_{i}{}^{k} 
F^{\beta  \gamma  ij} F_{\gamma  \varepsilon  jk} 
\nabla_{\alpha  }\Phi  \nabla^{\alpha  }\Phi  +   
 b_{20}   F^{\alpha  \beta  ij}  \nabla_{\alpha  
}F^{\epsilon  \varepsilon  }{}_{ik} \nabla_{\beta  
}F_{\epsilon  \varepsilon  j}{}^{k} \nn\\&\!\!\!\!\!\!\!\!&+   
 b_{21}   F_{\alpha  }{}^{\gamma  ij} F_{\beta  
}{}^{\varepsilon  }{}_{i}{}^{k} F_{\gamma  \varepsilon  jk} 
\nabla^{\alpha  }\Phi  \nabla^{\beta  }\Phi \! +  \! 
 b_{22}   F_{\alpha  }{}^{\gamma  ij} F_{\beta  
}{}^{\varepsilon  }{}_{i}{}^{k} F_{\gamma  \varepsilon  jk} 
\nabla^{\beta  }\nabla^{\alpha  }\Phi \! +\!   
 b_{23}   F_{\alpha  }{}^{\epsilon  }{}_{i}{}^{k} 
F^{\alpha  \beta  ij} F^{\nu  \gamma  
}{}_{jk} \nabla_{\epsilon  }H_{\beta  \nu  
\gamma  }\Big],\nn
\eeqa
where $b_1, b_2, \ldots, b_{23}$ represent the corresponding coupling constants. 
It includes single-trace terms, namely $\Tr(FFF)$ and $\Tr(FFFFF)$, as well as two-trace term  $\Tr(FF)\Tr(FFF)$. If one does not use field redefinition and only removes the redundancy due to integration by parts and various Bianchi identities, one would find the maximal basis, which has 31 couplings.

At the 7-derivative order, a comprehensive analysis reveals the existence of a total of 19,310 couplings. However, by eliminating redundancies resulting from field redefinitions, integration by parts, and the application of the Bianchi identity, it is possible to identify 1,288 independent couplings in the minimal basis. These independent couplings can be expressed in a particular scheme as follows:
\beqa
{\bf L}^{({5/2})}&\!\!\!\!=\!\!\!\!&\alpha'^2\sqrt{\alpha'}\Big[ \! c_{1}  
    F_{\alpha  }{}^{\gamma  kl} F^{\alpha  \beta  ij} 
F_{\beta  }{}^{\delta  }{}_{k}{}^{m} F_{\gamma  }{}^{\epsilon  
}{}_{m}{}^{o} F_{\delta  }{}^{\varepsilon  }{}_{o}{}^{n} 
F_{\epsilon  }{}^{\nu  }{}_{ln} F_{\varepsilon  \nu  ij}\!\!+\!\! 
    c_{2}  
    F_{\alpha  }{}^{\gamma  }{}_{i}{}^{k} F^{\alpha  \beta  
ij} F_{\beta  }{}^{\delta  lm} F_{\gamma  }{}^{\epsilon  o n} 
F_{\delta  }{}^{\varepsilon  }{}_{on} F_{\epsilon  }{}^{\nu  
}{}_{lm} F_{\varepsilon  \nu  jk} \nn\\&&+\cdots+
    c_{1287}  
    F^{\alpha  \beta  ij} F^{\gamma  \delta  }{}_{i}{}^{k} H_{
\alpha  \epsilon  \nu  } \nabla_{\varepsilon  }H_{\beta  
\gamma  \delta  } \nabla^{\nu  }F^{\epsilon  \varepsilon  
}{}_{jk}\!  + \!  c_{1288}  
    F_{\alpha  }{}^{\gamma  }{}_{i}{}^{k} F^{\alpha  \beta  
ij} F_{\beta  }{}^{\delta  }{}_{jk} \nabla_{\varepsilon  
}H_{\delta  \epsilon  \nu  } \nabla^{\nu  }H_{\gamma  
}{}^{\epsilon  \varepsilon  }\Big].\labell{L7}
 \eeqa
The expression above represents a subset of the independent couplings, with the ellipsis symbolizing an additional 1,284 terms that are not explicitly listed. These terms involve various combinations of $F$, $\nabla F$, $H$, $\nabla H$, $\nabla\Phi$, $\nabla\nabla\Phi$, and Riemann curvature. It includes single-trace terms, namely $\Tr(F^3)$, $\Tr(F^5)$, and $\Tr(F^7)$, as well as two-trace terms, namely $\Tr(F^2)\Tr(F^5)$ and $\Tr(F^3)\Tr(F^4)$. Additionally, there is a three-trace term $\Tr(F^2)\Tr(F^2)\Tr(F^3)$.
The parameters $c_1, c_2, \ldots, c_{1288}$ correspond to the respective coupling constants associated with these independent couplings. In this case, if one does not use field redefinition and only removes the redundancy due to integration by parts and various Bianchi identities, one would find the maximal basis, which has 1941 couplings.

The coupling constants in the above bases are background-independent and invariant under the YM gauge transformations. Hence, they may be fixed in a particular background which has one circle and in the particular gauge where the YM connection is zero but the YM field strength is not. In this background, the reduction of the couplings should be invariant under T-duality. Furthermore, in this YM gauge, the YM commutator terms are zero, which greatly simplifies the imposition of T-duality.
Once the coupling constants are fixed in this particular background and YM gauge, the background and gauge independence of the coupling constants guarantee that they are valid for any other background and any other gauges. 
In the next section, we will determine these coupling constants through the T-duality of the circular reduction, specifically within the chosen YM gauge.

\section{T-duality Constraint on the Minimal Bases}\label{sec.2}

The observation that the dimensional reduction of the classical effective action of string theory on a torus $T^{(d)}$ must be invariant under the $O(d,d, \mathbb{R})$ transformations \cite{Sen:1991zi,Hassan:1991mq,Hohm:2014sxa} indicates that the circular reduction of the couplings in the effective Lagrangian should be invariant under the discrete group $O(1,1, \mathbb{Z})$ or $\mathbb{Z}_2$-group, which consists only of non-geometrical transformations. Therefore, in order to impose this T-duality on the classical effective Lagrangian ${\bf L}_{\rm eff}$, we need to reduce the theory on a circle to obtain the $(D-1)$-dimensional effective Lagrangian $L_{\rm eff}(\psi)$, where $\psi$ collectively represents the base space fields. Then, we transform this  Lagrangian under the $\mathbb{Z}_2$-transformations to produce $L_{\rm eff}(\psi')$, where $\psi'$ is the T-duality transformation of $\psi$. The T-duality constraint on the effective Lagrangian is given by: 
\beqa
L_{\rm eff}(\psi)-L_{\rm eff}(\psi')=0\,.\labell{TL}
\eeqa
If we apply the T-duality once more to the above equation, we find $L_{\rm eff}[(\psi')']=L_{\rm eff}(\psi)$. This indicates that the T-duality transformation that satisfies the above relation must satisfy the $\mathbb{Z}_2$-group, i.e., $(\psi')'=\psi$.

To utilize the aforementioned constraint on the effective Lagrangian, one must first determine the basis in which  only the redundancies resulting from field redefinitions and Bianchi identities are eliminated. This basis comprises terms that are total derivative terms. Subsequently, the aforementioned constraint can be employed to determine the coupling constants in the bases as well as the T-duality transformations \cite{Garousi:2023qwj}.
However, to work with bases that have fewer couplings, it is necessary to remove the couplings in the bases that are total derivative terms, as demonstrated in finding the minimal bases in \reef{L5} and \reef{L7}. In this case, T-duality should be imposed on the effective action ${\bf S}_{\rm eff}$  which includes the minimal bases. Since some total derivative terms are eliminated from the Lagrangian in the constraint \reef{TL}, the remaining terms do not satisfy the  relation \reef{TL} unless certain total derivative terms are included in the base space to compensate for the absence of total derivative terms in the effective action. The T-duality constraint in this scenario is given by:
\beqa
S_{\rm eff}(\psi)-S_{\rm eff}(\psi')=\int d^{D-1}x \sqrt{-\bg}\nabla_a\Big[e^{-2\bphi}J^a(\psi)\Big],\labell{TS}
\eeqa
where $S_{\rm eff}(\psi)$ represents the circular reduction of the effective action, and $S_{\rm eff}(\psi')$ denotes its transformation under the T-duality transformation. In the above equation, $J^a$ is an arbitrary covariant vector composed of the $(D-1)$-dimensional base space fields. In this case, if we apply T-duality once more to \reef{TS}, we find $S_{\rm eff}[(\psi')']=S_{\rm eff}(\psi)$ up to certain total derivative terms in the base space. Therefore, once again, the T-duality transformation that satisfies the above relation must adhere to the $\mathbb{Z}_2$-group, i.e., $(\psi')'=\psi$. 

The constraint \reef{TS} can be employed to determine the coupling constants in the effective action ${\bf S}_{\rm eff}$ as well as the T-duality transformations. One approach is to impose the symmetry $(\psi')'=\psi$ on the T-duality transformation to reduce the parameters in the most general T-duality transformations. These reduced parameters can then be used in the above constraint to fix the coupling constants in the effective action, as well as the remaining parameters in the T-duality transformations \cite{Garousi:2019mca, Garousi:2023kxw, Garousi:2020gio}. 
Alternatively, one can use the most general T-duality transformations in the above constraint and solve the equation to determine the coupling constants of the effective action and the corresponding T-duality transformation. It should be noted that the resulting T-duality transformation must automatically satisfy the symmetry $(\psi')'=\psi$. 
In this paper, we adopt the latter approach.

In the context of the heterotic theory, it has been proposed that the truncated reduced action, which should be invariant under truncated T-duality transformations, can fix all coupling constants in the effective action ${\bf S}_{\rm eff}$ and in the truncated T-duality transformations \cite{Garousi:2024avb}. The proposed constraint is given by:
\beqa
S^L_{\rm eff}(\psi)-S^L_{\rm eff}(\psi^L)=\int d^{9}x \sqrt{-\bg}\nabla_a\Big[e^{-2\bphi}J^a(\psi)\Big]\,.\label{TSL}
\eeqa
In the above equation, $\psi^L$ represents the truncated T-duality transformation, which includes only the zeroth and first order terms of the scalar component of the YM gauge field in the base space. $S^L_{\rm eff}(\psi)$ also corresponds to the truncated reduction of the effective action. In the second term of the equation, only the zeroth and first order terms of the scalar should be retained.

The truncated T-duality transformation $\psi^L$ can be expanded in powers of $\alpha'$ as follows:
\beqa
\psi^L=\psi_0^L+\sum_{n=1}^{\infty}\frac{\alpha'^n}{n!}\Delta\psi^{L(n)}+\sum_{n=0}^{\infty}\alpha'^{n+\frac{1}{2}}\Delta\psi^{L(n+\frac{1}{2})}. \labell{gBuch}
\eeqa
Here, $n$ represents integer numbers and $\psi_0^L$ represents the truncated T-duality transformations at zeroth derivative, also known as the truncated Buscher rule. $\Delta\psi^{L(k)}$ for $k$ being integer and half-integer numbers represents their deformations at order $\alpha'^k$.
If the truncated circular reduction of the effective action and the vector $J^a$ have the following $\alpha'$-expansions:
\begin{equation}
 S^L_{\rm eff}=\sum^\infty_{k=0}\alpha'^kS^{L(k)},\quad  J^a=\sum^\infty_{k=0}\alpha'^kJ^a_{(k)}\,,
\end{equation}
and  $S^{L(k)}(\psi^L)$ has the following Taylor expansion  around the truncated Buscher rule $\psi_0^L$:
\begin{equation}
S^{L(k)}(\psi^L)=\sum^\infty_{m=0}\alpha'^mS^{L(k,m)}(\psi_0^L),
\end{equation}
where $m$ represents integer and half-integer numbers, then the $\mathbb{Z}_2$-constraint in \reef{TSL}  can be written as:
\beqa
\sum^\infty_{k=0}\alpha'^k\left[S^{L(k)}(\psi)-\sum^\infty_{m=0}\alpha'^{m}S^{L(k,m)}(\psi_0^L)-\int d^{9}x\, \partial_a\left[e^{-2\bphi}J^a_{(k)}(\psi)\right]\right]=0, \labell{TSn}
\eeqa
where $J^a_{(k)}$ is an arbitrary covariant vector composed of the $9$-dimensional base space fields at order $\alpha'^{k+1/2}$. In the above equation, we assume the base space is flat for simplicity, as the $\mathbb{Z}_2$-constraint for the coupling constants is the same for curved and flat base spaces, as observed in \cite{Garousi:2019mca}.
To determine the appropriate constraints on the effective actions, each term at every order of $\alpha'$ must be set to zero.

The T-duality constraint \reef{TSn} at order $\alpha'^0$ is given by
\begin{equation}
S^{L(0)}(\psi)-S^{L(0)}(\psi^L_0)- \int d^{9}x\, \partial_a\left[e^{-2\bphi}J_{(0)}^a(\psi)\right]=0, \labell{TS0}
\end{equation}
where $S^{L(0)}(\psi)$ is the circular reduction of the effective action at the two-derivative order, which includes the zeroth and the first order of the YM scalar. The effective action at the two-derivative order is \cite{Gross:1985rr,Gross:1986mw}
\beqa
{\bf S}^{(0)}&=&-\frac{2}{\kappa^2}\int d^{10}x \sqrt{-G}e^{-2\Phi}\Big[R-\frac{1}{12}H_{\alpha\beta\gamma}H^{\alpha\beta\gamma}+4\nabla_\alpha\Phi\nabla^\alpha\Phi-\frac{1}{4}F_{\mu\nu ij}F^{\mu\nu ij}\Big]\,.\labell{two2}
\eeqa
To perform the dimensional reduction of this action on a circle, we utilize the reduction scheme for the NS-NS and YM fields as proposed in \cite{Maharana:1992my, Bergshoeff:1995cg, Garousi:2024avb}:
 \beqa
G_{\mu\nu}=\left(\matrix{\bg_{ab}+e^{\varphi}g_{a }g_{b }& e^{\varphi}g_{a }&\cr e^{\varphi}g_{b }&e^{\varphi}&}\!\!\!\!\!\right),\, \, A_{\mu}{}^{ij}=\left(\matrix{\bar{A}_a{}^{ij}+e^{\varphi/2}\alpha^{ij}g_a&\cr e^{\varphi/2}\alpha^{ij}&}\!\!\!\!\!\right),\, \Phi=\bar{\phi}+\varphi/4\,,\labell{reduc}
\eeqa
\beqa
B_{\mu\nu}= \left(\matrix{\bb_{ab}+\frac{1}{2}(b_{a }g_{b }-b_{b }g_{a })+\frac{1}{2}e^{\varphi/2}\alpha_{ij}(g_a\bar{A}_b{}^{ij}-g_b\bar{A}_a{}^{ij})&b_{a }-\frac{1}{2}e^{\varphi/2}\alpha_{ij}\bar{A}_a{}^{ij}\cr - b_{b }+\frac{1}{2}e^{\varphi/2}\alpha_{ij}\bar{A}_a{}^{ij}&0&}\!\!\!\!\!\right),\nn
\eeqa
where $\bar{g}_{ab}$, $\bar{b}_{ab}$, $\bar{\phi}$, and $\bar{A}_a^{ij}$ represent the metric, B-field, dilaton, and YM gauge field in the base space, respectively. Additionally, $g_a$ and $b_b$ denote two vectors, while $\varphi$ and $\alpha^{ij}$ represent scalars within this space. By employing the above reduction scheme, one can derive the following reductions for different components of the $H$-field \cite{Maharana:1992my,Bergshoeff:1995cg}:
\beqa
H_{aby}&=&W_{ab}-\frac{1}{2}e^{\vp}\alpha^2 V_{ab}-e^{\vp/2}\alpha^{ij}\bF_{ab ij}\,,\nn\\
H_{abc}&=&\bH_{abc}+3g_{[a}W_{bc]}-\frac{3}{2}e^{\vp}\alpha^2g_{[a}V_{bc]}-3e^{\vp/2}\alpha_{ij}g_{[a}\bF_{bc]}{}^{ij}\,,
\eeqa
where $\alpha^2=\alpha^{ij}\alpha_{ij}$, $W_{ab}=\partial_a b_b-\partial_b b_a$, $V_{ab}=\partial_a g_b-\partial_b g_a$, $\bar{F}_{ab}{}^{ij}=\partial_a\bar{A}_b{}^{ij}-\partial_b\bar{A}_a{}^{ij}$, and $\bar{H}$ represents the torsion in the base space:
\beqa
\bH_{abc}&=&3\prt_{[a}\bb_{bc]}-\frac{3}{2}g_{[a}W_{bc]}-\frac{3}{2}b_{[a}V_{bc]}-\frac{3}{2}\bA_{[a}{}^{ij}\bF_{bc]}{}^{ij}\,.
\eeqa
Our notation for antisymmetry is such that, for example, $3g_{[a}W_{bc]}=g_aW_{bc}-g_bW_{ac}-g_cW_{ba}$. The torsion obeys the following Bianchi identity:
\begin{equation}
d\bar{H}+\frac{3}{2}V\wedge W+\frac{3}{4}\bF^{ij}\wedge \bF_{ij}=0\,. \labell{bian0}
\end{equation}
The truncated reduction of the action \reef{two2} up to certain total derivative terms, considering a flat base space, can be found in \cite{Garousi:2024avb}
\beqa
S^{L(0)}(\psi)&\!\!\!\!=\!\!\!\!&-\frac{2}{\kappa'^2}\int d^{9}x \,e^{-2\bphi}\Big[-\frac{1}{12}\bH^2+4\prt_a\bphi\prt^a\bphi-\frac{1}{4}\bF_{ab ij}\bF^{ab ij}\nn\\
&\!\!\!\!\!\!\!\!\!&-\frac{1}{4}\prt_a\vp\prt^a\vp-\frac{1}{4}e^{\vp}V^2-\frac{1}{4}e^{-\vp}W^2-\frac{1}{2}e^{\vp/2}\bF_{ab ij}V^{ab}\alpha^{ij}+\frac{1}{2}e^{-\vp/2}\bF_{ab ij}W^{ab}\alpha^{ij}\Big]\,,\labell{two3}
\eeqa
where $\kappa'$ is related to the 9-dimensional Newton's constant.  It has been observed in \cite{Garousi:2024avb} that the above action satisfies the constraint \reef{TS0} for the following truncated Buscher transformations $\psi_0^L$:
\beqa
 g_{a}\rightarrow b_{a}, b_{a}\rightarrow g_{a}, \vp\rightarrow -\vp, \alpha^{ij}\rightarrow -\alpha^{ij},\bA_a{}^{ij}\rightarrow \bA_a{}^{ij}, \bar{\eta}_{ab}\rightarrow \bar{\eta}_{ab},  \bb_{ab}\rightarrow \bb_{ab},\bphi\rightarrow \bphi\,,\labell{bucher}
 \eeqa
where $\bar{\eta}_{ab}$ is the flat metric of the base space. 

The constraint in \reef{TSn} at order $\sqrt{\alpha'}$ becomes
\begin{equation}
-S^{L(0,1/2)}(\psi_0^L)- \int d^{9}x\, \partial_a\left[e^{-2\bphi}J^a_{(1/2)}(\psi)\right]=S^{L(1/2)}(\psi_0^L)-S^{L(1/2)}(\psi). \label{TS3}
\end{equation}
Using the reductions in \reef{reduc}, one finds the  truncated reduction of the couplings in \reef{L3}:
\begin{equation}
S^{L(1/2)}(\psi)=-\frac{2}{\kappa'^2}\int d^{9}x \,e^{-2\bphi}\left[a_1 \bF_a{}^c{}_i{}^k\bF^{abij}\bF_{bcjk}+3a_1e^{\vp/2}\bF_{a}{}^c{}_i{}^k\bF_{bcjk}V^{ab}\alpha^{ij}\right]. \labell{two31}
\end{equation}
Its transformation under the truncated Buscher transformations \reef{bucher} is:
\begin{equation}
S^{L(1/2)}(\psi_0^L)=-\frac{2}{\kappa'^2}\int d^{9}x \,e^{-2\bphi}\left[a_1 \bF_a{}^c{}_i{}^k\bF^{abij}\bF_{bcjk}+3a_1e^{-\vp/2}\bF_{a}{}^c{}_i{}^k\bF_{bcjk}W^{ab}\alpha^{ij}\right]. \labell{two32}
\end{equation}
Since these terms are single-trace and include only field strengths $\bF, V, W$ without their derivatives, it is impossible to produce such terms by total derivative or by corrections to the leading-order T-duality transformation \reef{bucher}. Hence, the terms on the left-hand side of \reef{TS3} are zero, and the terms on the right-hand side must also be zero, which fixes the parameter to be zero. Therefore, T-duality dictates that
\begin{equation}
\Delta \psi^{L(\frac{1}{2})}=0, \quad a_1=0.
\end{equation}
The result for the effective action is consistent with the S-matrix method \cite{Gross:1986mw}.

Since $S^{(1/2)}=0$, its Taylor expansion is also zero, hence, $S^{(1/2,m)}=0$. Moreover, since $\Delta\psi^{L(1/2)}=0$, the Taylor expansion $S^{(1,1/2)}$ is also zero. As a result, the constraint in \reef{TSn}  at order $\alpha'\sqrt{\alpha'}$ becomes:
\begin{equation}
-S^{L(0,3/2)}(\psi_0^L)- \int d^{9}x\, \partial_a\left[e^{-2\bphi}J^a_{(3/2)}(\psi)\right]=S^{L(3/2)}(\psi_0^L)-S^{L(3/2)}(\psi). \labell{TS31}
\end{equation}
Using the reductions in \reef{reduc}, it is straightforward to find the  truncated reduction of the couplings in \reef{L5} to generate $S^{L(3/2)}(\psi)$. Then, by applying the truncated Buscher transformations in \reef{bucher}, one can calculate its corresponding $S^{L(3/2)}(\psi_0^L)$. It is important to note that the right-hand side is odd under the truncated Buscher transformation. Hence, $S^{L(0,3/2)}(\psi_0^L)$ must also be odd, up to certain total derivative terms.

To determine the first term on the left-hand side of the above equation, we require the inclusion of 3-derivative corrections to the truncated Buscher rules \reef{bucher}. In other words, we need to consider the expanded form of the Buscher transformations that includes terms involving three derivatives.
\beqa
&&\varphi^L= -\varphi+\alpha'\sqrt{\alpha'}\Delta\varphi^{(3/2)}(\psi)+\cdots
\,,\quad g^L_{a}= b_{a}+\alpha'\sqrt{\alpha'}e^{\varphi/2}\Delta g^{(3/2)}_a(\psi)+\cdots
\,,\quad\nn\\
&&b^L_{a}= g_{a}+\alpha'\sqrt{\alpha'}e^{-\varphi/2}\Delta b^{(3/2)}_a(\psi)+\cdots
\,,\quad \bar{g}_{ab}^L=\eta_{ab}+\alpha'\sqrt{\alpha'}\Delta \bar{g}^{(3/2)}_{ab}(\psi)+\cdots
\,,\quad\labell{T22}\\
&&\bar{H}_{abc}^L=\bar{H}_{abc}+\alpha'\sqrt{\alpha'}\Delta\bar{H}^{(3/2)}_{abc}(\psi)+\cdots
\,,\quad\bar{\phi}^L= \bar{\phi}+\alpha'\sqrt{\alpha'}\Delta\bar{\phi}^{(3/2)}(\psi)+\cdots\,,\nn\\
&&(\bar{A}_a^L)^{ij}=\bar{A}_a{}^{ij}+\alpha' \sqrt{\alpha'}\Delta\bar{A}_a^{(3/2)}{}^{ij}(\psi)+\cdots\,,\quad(\alpha^L)^{ij}=-\alpha^{ij}+\alpha' \sqrt{\alpha'}\Delta\alpha^{(3/2)}{}^{ij}(\psi)+\cdots.\nn
\eeqa
Note that the corrections also contain non-zero terms at the 2-derivative order, denoted as $\Delta\psi^{L(1)}$, which have been derived in \cite{Garousi:2024avb}. However, since we have $\Delta\psi^{L(\frac{1}{2})}=0$, these corrections do not appear in the Taylor expansion $S^{L(0,3/2)}(\psi_0^L)$ because  $\Delta\psi^{L(1)}\Delta\psi^{L(\frac{1}{2})}=0$, hence, they  play no role in studying the T-duality of the 5-derivative couplings.

The T-duality transformed base space fields $\psi^L$ must satisfy the Bianchi identity \reef{bian0}. Therefore, the correction $\Delta\bar{H}^{(3/2)}_{abc}$ is related to the corrections $\Delta b^{(3/2)}_a$, $\Delta \bar{g}^{(3/2)}_{ab}$, $\Delta\bar{A}_a^{(3/2)}{}^{ij}$, and an arbitrary two-form $\Delta B_{ab}^{(3/2)}$, all of which are at the 3-derivative order. This relation is given by \cite{Garousi:2024avb}:
\begin{equation}
\Delta \bar{H}^{(3/2)}_{abc}=3\partial_{[a}\Delta B^{(3/2)}_{bc]}-3e^{\varphi/2}V_{[ab}\Delta g^{(3/2)}_{c]}-3e^{-\varphi/2}W_{[ab}\Delta b^{(3/2)}_{c]}-3\bar{F}_{[ab}{}^{ij}\Delta \bar{A}^{(3/2)}_{c]ij}.\labell{rel2}
\end{equation}
Here, $\Delta B^{(3/2)}_{ab}$ represents an arbitrary two-form at the 3-derivative order. It is important to note that the truncated T-duality transformations \reef{T22} must satisfy the $\mathbb{Z}_2$-group. Consequently, the deformations at the order of $\alpha'\sqrt{\alpha'}$ need to adhere to the following constraint:
\beqa
-\Delta\varphi^{(3/2)}(\psi)+\Delta\varphi^{(3/2)}(\psi_0^L) =0 &;&
\Delta b_a^{(3/2)}(\psi)+\Delta g_a^{(3/2)}(\psi_0^L) =0\,,\nn\\
\Delta g_a^{(3/2)}(\psi)+\Delta b_a^{(3/2)}(\psi_0^L)=0 &;&
\Delta \bar{g}_{ab}^{(3/2)}(\psi)+\Delta \bar{g}_{ab}^{(3/2)}(\psi_0^L)=0\,,\nn\\
\Delta\bar{\phi}^{(3/2)}(\psi)+\Delta\bar{\phi}^{(3/2)}(\psi_0^L) =0 &;&
\Delta B_{ab}^{(3/2)}(\psi)+\Delta B_{ab}^{(3/2)}(\psi_0^L)=0\,,\nn\\
\Delta \bar{A}_a^{(3/2)ij}(\psi)+\Delta \bar{A}_a^{(3/2)ij}(\psi_0^L)=0 &;&
-\Delta \alpha^{(3/2)ij}(\psi)+\Delta \alpha^{(3/2)ij}(\psi_0^L)=0\,.\label{Z22}
\eeqa
The corrections need to be constructed by considering all possible contractions of $\partial \varphi$, $\partial \bar{\varphi}$, $e^{\varphi/2}V$, $e^{-\varphi/2}W$, $\bar{H}$, $\bar{F}_{ab}{}^{ij}$, and their derivatives at the order of $\alpha'\sqrt{\alpha'}$, with arbitrary coefficients. These corrections should include terms at both zeroth and first order in the scalar $\alpha^{ij}$. One approach is to impose the above constraint  on these corrections, which would reduce the number of parameters involved. These modified corrections can then be used in the constraint \reef{TS31}. Alternatively, one can directly insert the most general form of the corrections into the constraint \reef{TS31}. In this case, the constraint \reef{TS31} would determine the coupling constants in the effective action and also fix the parameters in the T-duality transformations, such that they satisfy the relations mentioned above.

By employing the T-duality transformations \reef{T22} and the relation \reef{rel2}, one can derive the Taylor expansion of the T-duality transformation of the truncated reduced action at the leading order \reef{two3}, centered around the truncated Buscher transformations $\psi_0^L$. The resulting expression is as follows:
\beqa
 S^{L(0,3/2)}(\psi_0^L)
 \!\!\!\!\!\!\!\!\!\!&&=\! -\frac{2\alpha'^{3/2}}{\kappa'^2}\int \!\!d^{9}x e^{-2\bphi} \,  \Big[ \!\Big(\!\frac{1}{4}\prt^a\vp\prt^b\vp\!-\!2\prt^a\prt^b\bphi\!+\!\frac{1}{4}\bH^{acd}\bH^b{}_{cd} \!+\!\frac{1}{2}(e^\vp V^{ac}V^b{}_{c}\!+\!e^{-\vp} W^{ac}W^b{}_{c})\nn\\
 &&+\bF^{acij}\bF^b{}_{cij}+(e^{\vp/2}V^{ac}-e^{-\vp/2}W^{ac})\bF^b{}_{cij}\alpha^{ij}\Big)\Delta\bar{g}_{ab}^{(3/2)}+\Big(2\prt_c\prt^c\bphi-2\prt_c\bphi\prt^c\bphi -\frac{1}{24}\bH^2\nn\\&\!\!\!\!\!\!\!&-\frac{1}{8}\prt_c\vp\prt^c\vp-\frac{1}{8}(e^{\vp} V^2+e^{-\vp}W^2)+\frac{1}{2}\bF_{abij}\bF^{abij}+(e^{\vp/2}V^{ab}-e^{-\vp/2}W^{ab})\bF_{abij}\alpha^{ij}\Big)\nn\\&\!\!\!\!\!\!\!&\times(\eta^{ab}\Delta\bar{g}_{ab}^{(3/2)}-4\Delta\bphi^{(3/2)})-\Big(\frac{1}{2}\prt_a\prt^a\vp-\prt_a\bphi\prt^a\vp+\frac{1}{4}(e^{\vp/2}V^{ab}+e^{-\vp/2}W^{ab})\bF_{abij}\alpha^{ij}\nn\\&\!\!\!\!\!\!\!&-\frac{1}{4}(e^\vp V^2-e^{-\vp}W^2)\Big)\Delta\vp^{(3/2)}+\Big(2e^{-\vp/2}\prt_b\bphi W^{ab}- e^{-\vp/2}\prt_b W^{ab}+e^{-\vp/2}\prt_b\vp W^{ab}\nn\\&&+\frac{1}{2}e^{\vp/2}\bH^{abc}V_{bc}-2(\prt_b\bphi+\frac{1}{4}\prt_b\vp)\bF^{abij}\alpha_{ij}+\prt_b\bF^{abij}\alpha_{ij}\Big)\Delta g_a^{(3/2)} +\Big(2e^{\vp/2}\prt_b\bphi V^{ab}\nn\\&&- e^{\vp/2}\prt_b V^{ab}-e^{\vp/2}\prt_b\vp V^{ab}+\frac{1}{2}e^{-\vp/2}\bH^{abc}W_{bc}+2(\prt_b\bphi-\frac{1}{4}\prt_b\vp)\bF^{abij}\alpha_{ij}\nn\\&\!\!\!\!\!\!\!&-\prt_b\bF^{abij}\alpha_{ij}\Big)\Delta b_a^{(3/2)}+\Big(\frac{1}{2}\prt_a\bH^{abc}-\bH^{abc}\prt_a\bphi\Big)\Delta B_{bc}^{(3/2)}+\Big(\frac{1}{2}\bH^{abc}\bF_{bcij}-\prt_b\bF^{ab}{}_{ij}\nn\\&\!\!\!\!\!\!\!&+2\prt_b\bphi\bF^{ab}{}_{ij}+\alpha_{ij}[2(\prt_b\bphi-\frac{1}{4}\prt_b\vp)e^{\vp/2}V^{ab}-2(\prt_b\bphi+\frac{1}{4}\prt_b\vp)e^{-\vp/2}W^{ab}-e^{\vp/2}\prt_b V^{ab}\nn\\&\!\!\!\!\!\!\!&+e^{-\vp/2}\prt_bW^{ab}]\Big)\Delta \bA_a^{(3/2)ij}+\frac{1}{2}\bF_{abij}(e^{\vp/2}V^{ab}-e^{-\vp/2}W^{ab})\Delta \alpha^{(3/2)ij}\Big] \,,\labell{d3S}
 \eeqa
where we have also removed certain total derivative terms. It is important to emphasize that the terms presented above exhibit an odd behavior under the truncated Buscher transformations. Assuming that the correction $\Delta \alpha^{(3/2)ij}$ includes terms at the zeroth order of $\alpha^{ij},$ it becomes necessary to consider the terms in the reduced action that involve the second order of $\alpha^{ij}$. Consequently, this consideration introduces terms in the equation above that contain $\alpha_{ij} \Delta \alpha^{(3/2)ij}$. However, when multiplied by the appropriate terms resulting from the corresponding term in the reduced action, this contribution fails to maintain the desired odd behavior under the Buscher transformations (see the explicit form of this term in \cite{Garousi:2024avb}). Therefore, it can be concluded that $\Delta \alpha^{(3/2)ij}$ cannot contain terms at the zeroth order of $\alpha^{ij}$.

The vector $J^a_{(3/2)}$ in the total derivative term of the T-duality constraint \reef{TS31} should be constructed by considering contractions of $\partial \varphi$, $\partial \bar{\varphi}$, $e^{\varphi/2}V$, $e^{-\varphi/2}W$, $\bar{H}$, and $\bar{F}_{ab}{}^{ij}$ at the 4-derivative order, with arbitrary coefficients. These constructions should also include terms at both the zeroth and first order in the scalar $\alpha^{ij}$.

 In order to solve the equation \reef{TS31}, it is crucial to impose the Bianchi identities associated with the field strengths $\bH$, $\bF$, $V$, and $W$. We impose the $\bH$-Bianchi identity in its gauge-invariant form, while for the other Bianchi identities, we impose them in a non-gauge-invariant form.
To enforce the $\bH$-Bianchi identity, we introduce the gauge-invariant tensor $\mathbf{H}_{abcd}$, which is totally antisymmetric and of order $\alpha'$. We then consider all possible scalar contractions involving $\partial \varphi$, $\partial \bar{\varphi}$, $e^{\varphi/2}V$, $e^{-\varphi/2}W$, $\bar{H}$, $\mathbf{H}$, $\bF^{ij}$, and their derivatives up to the 5-derivative order, which linearly include the tensor $\mathbf{H}$. These contractions should include terms at both the zeroth and first order in the scalar $\alpha^{ij}$, and the coefficients of these gauge-invariant terms are arbitrary.
Next, we incorporate the following gauge-invariant relation for the newly defined tensor:
\beqa
{\bf H}_{abcd} &= \partial_{[a} \bar{H}_{bcd]} +\frac{3}{2} V_{[ab}W_{cd]}+\frac{3}{4} \bF_{[ab}{}^{ij}\bF_{cd]}{}_{ij}  \labell{HWV}\,.
\eeqa
We denote the resulting gauge-invariant scalar as ${\bf BI}(\psi)$. This scalar is then added to \reef{TS31}, thereby imposing the $\bH$-Bianchi identity in a gauge-invariant form.
To impose the Bianchi identities corresponding to $\bF$, $V$, and $W$, whenever there are terms with derivatives of these field strengths, we replace them with their corresponding potentials.
As a result, equation \reef{TS31} can be expressed in terms of independent but non-gauge-invariant couplings. The coefficients of these independent terms must be set to zero, resulting in a system of linear algebraic equations involving all the parameters.

Upon solving the system of algebraic equations, we discover two distinct sets of solutions for the parameters involved. The first set represents relationships exclusively among the parameters in the deformations of the truncated Buscher rules, the parameters corresponding to the $\bH$-Bianchi identity, and the parameters in the total derivative terms. These solutions satisfy the homogeneous equation corresponding to \reef{TS31}, with the right-hand side being zero. However, this general solution is not of interest to us.

The other solution, which is the particular solution of the non-homogeneous equation \reef{TS31} and the one of interest, provides an expression for these parameters in terms of the coupling constants $b_1, b_2, \ldots$. This solution also establishes the relationships between the coupling constants. In our analysis, we will consider this particular solution that sets all parameters in the effective action and their corresponding corrections to the Buscher rules to zero, i.e.,
\beqa
\Delta\psi^{L(\frac{3}{2})}=0\,, && b_1=b_2=\cdots=b_{23}=0\,.
\eeqa
In fact, there are 12 couplings with coefficients $b_1, b_5, b_6, b_7, b_8, b_9, b_{11}, b_{12}, b_{13}, b_{17}, b_{18}, b_{20}$ that appear solely in the algebraic equation, without any contribution from total derivatives, corrections to the Buscher transformations, or the Bianchi identities. As a result, they are constrained to be zero. The other coefficients appear in the algebraic equation in combination with the other parameters of the total derivative terms, corrections to the Buscher transformations, and the Bianchi identities. However, the algebraic equation fixes them to zero as well. Therefore, the effective action at the 5-derivative order in the particular gauges where the YM potential is zero but its field strength is not zero, vanishes identically. The YM gauge symmetry then dictates that the effective action is zero in any other gauges as well.

Given that $S^{L(1/2)},\,S^{L(3/2)},\, \Delta\psi^{L(\frac{1}{2})}$ and $\Delta\psi^{L(\frac{3}{2})}$ are zero,  the constraint in \reef{TSn} at the order $\alpha'^2\sqrt{\alpha'}$ takes the form:
\beqa
-S^{L(0,5/2)}(\psi_0^L)- \int d^{9}x\, \partial_a\Big[e^{-2\bphi}J^a_{(5/2)}(\psi)\Big]=S^{L(5/2)}(\psi_0^L)-S^{L(5/2)}(\psi).\labell{TS32}
\eeqa
By employing the circular reductions in \reef{reduc}, one can compute the  truncated reduction of the couplings in \reef{L7} to generate $S^{L(5/2)}(\psi)$. Subsequently, by utilizing the truncated Buscher transformations in \reef{bucher}, one can calculate the corresponding $S^{L(5/2)}(\psi_0^L)$.

To determine the first term on the left-hand side of the equation \reef{TS32}, we consider the 5-derivative corrections to the truncated Buscher rules \reef{bucher}, given by:
\beqa
&&\varphi^L= -\varphi+\alpha'^2\sqrt{\alpha'}\Delta\varphi^{(5/2)}(\psi)+\cdots
\,,\quad g^L_{a}= b_{a}+\alpha'^2\sqrt{\alpha'}e^{\varphi/2}\Delta g^{(5/2)}_a(\psi)+\cdots
\,,\quad\nn\\
&&b^L_{a}= g_{a}+\alpha'^2\sqrt{\alpha'}e^{-\varphi/2}\Delta b^{(5/2)}_a(\psi)+\cdots
\,,\quad \bar{g}_{ab}^L=\eta_{ab}+\alpha'^2\sqrt{\alpha'}\Delta \bar{g}^{(5/2)}_{ab}(\psi)+\cdots
\,,\quad\labell{T23}\\
&&\bar{H}_{abc}^L=\bar{H}_{abc}+\alpha'^2\sqrt{\alpha'}\Delta\bar{H}^{(5/2)}_{abc}(\psi)+\cdots
\,,\quad\bar{\phi}^L= \bar{\phi}+\alpha'^2\sqrt{\alpha'}\Delta\bar{\phi}^{(5/2)}(\psi)+\cdots\,,\nn\\
&&(\bar{A}_a^L)^{ij}=\bar{A}_a{}^{ij}+\alpha'^2\sqrt{\alpha'}\Delta\bar{A}_a^{(5/2)}{}^{ij}(\psi)+\cdots\,,\quad(\alpha^L)^{ij}=-\alpha^{ij}+\alpha'^2\sqrt{\alpha'}\Delta\alpha^{(5/2)}{}^{ij}(\psi)+\cdots.\nn
\eeqa
It is worth noting that these corrections also contain terms at the 2-derivative and 4-derivative orders, i.e., $\Delta\psi^{L(1)}$ and $\Delta\psi^{L(2)}$. However, since $\Delta\psi^{L(\frac{1}{2})}=0=\Delta\psi^{L(\frac{3}{2})}=0$, these corrections do not play a role in studying the T-duality of the 7-derivative couplings. Thus, the first term on the left-hand side of the constraint \reef{TS32} can be obtained by replacing the 3-derivative correction of the Buscher rules in \reef{d3S} with the 5-derivative corrections. The subsequent steps follow a similar procedure. The result we obtain is as follows:
\beqa
\Delta\psi^{L(\frac{5}{2})}=0\,, && c_1=c_2=\cdots=c_{1288}=0.
\eeqa
Hence, the effective action at the 7-derivative order is zero.

Given that $S^{L(1/2)},\,S^{L(3/2)},\,S^{L(5/2)},\,\Delta\psi^{L(\frac{1}{2})},\,\Delta\psi^{L(\frac{3}{2})}$, and $\Delta\psi^{L(\frac{5}{2})}$ are zero, the constraint in \reef{TSn} at the order $\alpha'^3\sqrt{\alpha'}$ can be expressed as follows:
\beqa
-S^{L(0,7/2)}(\psi_0^L)- \int d^{9}x\, \partial_a\Big[e^{-2\bphi}J^a_{(7/2)}(\psi)\Big]=S^{L(7/2)}(\psi_0^L)-S^{L(7/2)}(\psi).\labell{TS33}
\eeqa
However, performing explicit calculations at this order and beyond is extremely challenging for computers due to the extensive length of the calculations involved. While there are no fundamental obstacles in performing these calculations, the main difficulty lies in the extensive computational effort required. We expect that the same results we have obtained for the 3-, 5-, and 7-derivative cases hold true for all higher odd-derivative couplings, namely:
\beqa
\Delta\psi^{L(n+\frac{1}{2})}=0\,, && S^{L(n+\frac{1}{2})}=0\,;\,\,\, n=0,1,2,\cdots\,.
\eeqa
Hence, the truncated T-duality transformation \reef{gBuch} can only have even-derivative contributions, i.e., 
\beqa
\psi^L=\psi_0^L+\sum_{n=1}^{\infty}\frac{\alpha'^n}{n!}\Delta\psi^{L(n)}. \labell{gBuch1}
\eeqa
Furthermore, the effective action also consists solely of even-derivative gauge invariant couplings, i.e.,
\begin{equation}
 \bS_{\rm eff}=\sum^\infty_{n=0}\alpha'^n\bS^{(n)}\,.
\end{equation}
However, it is important to note that this result does not imply the absence of contact terms in the sphere-level S-matrix element of the heterotic theory that involve an odd number of momenta. In fact, these contact terms should be accounted for by the non-zero commutator terms present in the even-derivative couplings of field strengths. For example, consider the three-point function of three YM vertex operators, which includes one momentum \cite{Gross:1986mw}. This particular configuration is indeed reproduced by the commutator term present in the leading-order action \reef{two2}. Thus, while the odd-derivative couplings vanish, the contact terms involving odd numbers of momenta are still captured by the commutator terms within the even-derivative couplings.

\section{Discussion}

The classical effective action of string theory, when subjected to circular reduction, exhibits the $O(1,1,\MR)$ symmetry, as discussed in \cite{Sen:1991zi,Hassan:1991mq,Hohm:2014sxa}. The determination of coupling constants within the effective action can be achieved by applying the non-geometric subgroup $O(1,1,\MZ)$ or the associated generalized Buscher transformations to the gauge invariant independent couplings at any order of derivatives.
In the context of heterotic string theory, the circular reduction of the effective action and the generalized Buscher transformations demonstrate nonlinearity with respect to the scalar component of the YM gauge field. A recent paper  \cite{Garousi:2024avb} has introduced a truncated approach, where the reduced action and generalized Buscher transformations treat the scalar field as a constant and retain only the zeroth and first orders of the scalar field. By employing this truncated version of the generalized Buscher transformations, the coupling constants within the effective action can still be determined.
In this study, we have applied this constraint to the gauge invariant independent couplings with an odd number of derivatives. Specifically, we have provided explicit demonstrations that for 3-, 5-, and 7-derivative couplings, the odd-derivative terms vanish under this imposed constraint. We further speculate that this result holds true for all higher odd-derivative couplings as well. This is in contrast to the couplings in type I superstring theory, where non-zero couplings of, for example, 5 YM field strength are present \cite{Kitazawa:1987xj}.

We have applied T-duality to the minimal bases  \reef{L5} and \reef{L7} in order to determine their coupling constants. This is based on the assumption that the combinations of couplings, which can be eliminated through field redefinition, integration by parts, and Bianchi identities, may also be invariant under T-duality. This assumption is valid for the NS-NS couplings \cite{Garousi:2019mca, Garousi:2023kxw, Garousi:2020gio}, which may be a consequence of the fact that the Buscher transformations for the NS-NS fields in the reduction scheme \cite{Maharana:1992my} are linear.
However, this assumption may not be valid when considering YM fields because the corresponding Buscher transformations in the presence of YM field are nonlinear \cite{Bergshoeff:1995cg}. This assumption is not valid even for truncated Buscher transformations. To see this, consider, for instance, the following examples of such couplings with an overall factor $a$:
\beqa
a\Big[\frac{1}{2} F_{\alpha }{}^{\gamma }{}_{i}{}^{k} 
F^{\alpha \beta i j} F^{\epsilon \varepsilon 
}{}_{j k} H_{\beta \gamma }{}^{\mu } H_{\epsilon 
\varepsilon \mu } - 2 F_{\alpha }{}^{\beta i j} 
F_{\gamma }{}^{\varepsilon }{}_{j k} F^{\gamma 
\epsilon }{}_{i}{}^{k} H_{\beta \epsilon 
\varepsilon } \nabla^{\alpha }\Phi + F_{\alpha }{}^{\gamma 
}{}_{i}{}^{k} F^{\alpha \beta i j} 
H_{\beta \gamma \varepsilon } \nabla_{\epsilon }F^{\epsilon 
\varepsilon }{}_{j k}\Big]\labell{multi}
\eeqa
 These couplings can be removed by a field redefinition of the form $A_{\alpha ij}=A_{\alpha ij}-F_\beta{}^\mu {}_{jk}F^{\beta\nu}{}_i{}^k H_{\alpha\nu\mu}$, along with certain total derivative terms. However, the reduction of these couplings minus their transformation under the Buscher rules \reef{bucher} is a complicated expression that cannot be removed through any corrections to the Buscher rules. This can be seen by considering the reduction minus T-duality for the case where $\nabla \bF=\nabla V=\nabla W=\nabla\bH=W=\nabla \bphi=\nabla\nabla \bphi=0$, which is 
\beqa
&&a\Big[- \frac{1}{2} e^{\vp} \bF_{ab}{}^{ij} 
\bF_{c}{}^{e}{}_{i}{}^{k} \bF_{dejk} V^{ab} V^{cd} + e^{\vp/2} \bF_{c}{}^{e}{}_{jk} \bF^{cd}{}_{i}{}^{k} \nabla^a \vp \bH_{bde} 
V_{a}{}^{b} \alpha^{ij}\nn\\&& -  \frac{1}{2} e^{ \vp/2} 
\bF_{a}{}^{ckl} \bF_{bck}{}^{m} \bF_{delm} \bF^{de}{}_{ij} V^{ab} 
\alpha^{ij}
 + \frac{1}{2} e^{ \vp/2} 
\bF_{c}{}^{e}{}_{jk} \bF^{cd}{}_{i}{}^{k} \bH_{ab}{}^{f} \bH_{def} 
V^{ab} \alpha^{ij}\nn\\&& + e^{ \vp/2} \bF_{a}{}^{c}{}_{i}{}^{k} 
\bF^{de}{}_{jk} \bH_{bc}{}^{f} \bH_{def} V^{ab} \alpha^{ij} -  
\frac{1}{4} e^{ \vp/2} \bF_{b}{}^{d}{}_{i}{}^{k} \bF_{cdjk} 
\nabla_a\vp  \nabla^a\vp V^{bc} \alpha^{ij} \nn\\&&-  \frac{1}{2} 
e^{ \vp/2} \bF_{b}{}^{d}{}_{i}{}^{k} \bF_{cdjk} \nabla_a\nabla^a\vp
V^{bc} \alpha^{ij} -  \frac{1}{2} e^{ \vp/2} 
\bF_{abi}{}^{k} \bF^{de}{}_{jk} \nabla^a\vp \bH_{cde} V^{bc} 
\alpha^{ij}\Big]
\eeqa
One can verify that these terms cannot be removed by any corrections to the Buscher rules. If the T-duality on the effective action fixes the coefficient $a$ to be non-zero, then one is not allowed to remove the couplings \reef{multi} from the list of couplings in the basis.

The observed phenomenon suggests that in the presence of YM fields, applying T-duality to the minimal basis, where field redefinitions are imposed, may not correctly determine the coupling constants. The correct approach to enforce T-duality and fix the coupling constants is by imposing it on the maximal basis, where field redefinitions are not imposed. However, in this case, T-duality is unable to fix all coupling constants, as there are several  parameters present in the resulting T-duality invariant action that can subsequently be eliminated through field redefinition. 
If the number of constraints that T-duality produces in the minimal basis is the same as the number of constraints it produces in the maximal basis, then there are no sets of couplings with the above property.
At the four-derivative order, there are no such couplings because the number of constraints between the coupling constants for the maximal basis is 24, which is equal to the number of constraints between the coupling constants in the minimal basis \cite{Garousi:2024avb}.
At the five-derivative order, we have discovered that applying T-duality to the minimal basis produces 23 relations, which  matches the number of minimal couplings. However, when applying T-duality to the maximal basis, it yields 25 relations, while there are a total of 31 couplings. This implies the existence of two sets of couplings that satisfy the aforementioned property. 
By incorporating the 25 relations from the maximal basis T-duality constraint, we verified that the remaining parameters in the action can be eliminated through field redefinition\footnote{ This calculation can be done by using the equation \reef{DLK} in the Appendix, in which $\Delta= L$, where $L$ is the Lagrangian that has 6 parameters. This Lagrangian is the result of inserting the 25 relations in the maximal basis. When one solves this equation, one finds that the parameters in field redefinitions and total derivative terms can be written in terms of the 6 parameters in $L$.
In other words, the equation \reef{DLK} does not produce any relation between the 6 parameters. This means the 6 parameters can be removed by field redefinitions and integration by parts. }.
Consequently, there are no physical couplings in the maximal basis at the five-derivative order.
It should be noted that the above property exists even for the abelian case, where the YM gauge field has no internal indices. In fact, we have performed the calculation for the abelian case at the five-derivative order and found 18 couplings in the minimal basis and 54 couplings in the maximal basis. The T-duality produces 18 relations in the minimal basis, whereas it produces 42 relations in the maximal basis. Again, when one inserts these 42 relations into the maximal basis, the remaining terms can be removed by field redefinitions\footnote{At the three-derivative order in the abelian case, there is only one independent coupling: $a F^{\beta\gamma}H_{\alpha\beta\gamma}\nabla^\alpha\Phi$. The T-duality fixes the coefficient of this coupling to be zero.}.

Furthermore, we imposed T-duality on the maximal basis at the seven-derivative order and found that it produces 1379 relations among the 1941 couplings. In contrast, T-duality on the minimal basis, which has 1288 couplings, generates 1288 relations. Hence, there are 91 sets of couplings that exhibit the mentioned property. After incorporating these 1379 relations into the action within the maximal scheme, we observed that all the remaining parameters in the resulting action can be removed through field redefinition. Therefore, there are no physical couplings at the seven-derivative order either.

The T-duality constraints \reef{TS31}, \reef{TS32}, \reef{TS33} have similar patterns and similar solutions, where the corresponding effective actions are zero. One may try to prove by induction that such a pattern dictates the vanishing of the coupling constant. 
However, there are similar patterns for the couplings at the four-derivative order, which produce  non-zero coupling constants \cite{Garousi:2024avb}. At the six-derivative order, the T-duality constraint has a different pattern that includes the Taylor expansion of the four-derivative couplings as well. This causes the coupling constant to be non-vanishing. If one removes this extra term, then the T-duality would find  a similar pattern as in \reef{TS31}, which produces the wrong result of   vanishing coupling constants.
At the eight-derivative order, the T-duality also produces extra terms from the Taylor expansion of the 4- and 6-derivative orders. If one removes such terms, then the T-duality would find  a similar pattern as in \reef{TS31}. However, using the fact that at the 8-derivative order, there are couplings with the coefficient $\z(3)$ which are not present in the 4- and 6-derivative orders, the simplified pattern should produce  non-vanishing coupling constants.
Hence, it seems it is non-trivial to prove from the pattern  \reef{TS31}, \reef{TS32}, \reef{TS33}  that the odd-derivative couplings are zero, without using explicit calculations.

We have found that the T-duality of the circular reduction of the effective action constrains all odd-derivative gauge-invariant couplings to be zero. This result is consistent with the $O(d,d+16)$ symmetry of the cosmological reduction of the effective action \cite{Hohm:2014sxa}. 
For zero YM field, it has been observed in \cite{Hohm:2015doa,Hohm:2019jgu} that the cosmological reduction of the effective action can be written in a specific scheme in terms of the trace of various orders of the first time derivative of the $O(d,d)$ matrix $\cS$. On the other hand, the trace of an odd number of $\dot{\cS}$ is zero \cite{Hohm:2015doa,Hohm:2019jgu}.
In the presence of a YM field, the $O(d,d)$ matrix $\cS$ is extended to an $O(d,d+16)$ matrix \cite{Maharana:1992my,Hohm:2014sxa}. In the specific gauge used in this paper that removes all the commutator terms, the cosmological reduction of the couplings should have $O(d,d+16)$ symmetry. They should be written in terms of the trace of various orders of $\dot{\cS}$, where $\cS$ now includes NS-NS and YM fields. Then, the observation that the trace of an odd number of $\dot{\cS}$ is zero dictates that there are no odd-derivative couplings in the heterotic theory.

In a recent paper \cite{Garousi:2024rzh}, it was observed that T-duality fixes the coupling constants in the NS-NS bases that have an odd number of $B$-fields to be zero, both in bosonic and superstring theories. This finding, along with the results presented in this paper, suggests that T-duality has the ability to determine the couplings in a basis up to an overall factor, as long as that factor is non-zero within the context of string theory. In other words, if a coupling is zero in string theory, T-duality will yield that result. However, if the overall factor of the couplings in a basis is non-zero in any of the string theories and their values depend on the specific type of string theory, then T-duality cannot fix it, since T-duality is a symmetry of all string theories.

It has been observed in \cite{Garousi:2024avb} that T-duality imposes constraints on the single-trace couplings of the YM field strength at the $n$-derivative order, with the exception of $n=2$, resulting in their vanishing. In the present paper, we have made a similar finding: the couplings involving single-trace terms with an odd number of YM field strengths, multiplied by other NS-NS terms or $\Tr(FF)$, are likewise forced to be zero. For instance, we have determined that the coupling $\Tr(FF)\Tr(FFF)$ is zero. This observation suggests that similar constraints may apply to even-derivative gauge invariant couplings as well. If this hypothesis holds true, one would expect the YM field strength and its derivatives to appear in the even-derivative couplings in the form of different orders of $\Tr(FF)$. Performing calculations for the 6-derivative and 8-derivative orders would be of great interest in order to verify this speculation.

We have observed how specifically fixing the YM gauge symmetry facilitates the identification of independent covariant couplings at various orders of derivatives.
In both bosonic and superstring theory, the nonabelian YM field arises on stacks of D$_p$-branes. One might attempt to employ a similar method to identify the independent covariant YM couplings for D$_p$-branes. However, in this case, it is not clear how to impose world-volume diffeomorphism symmetry for the nonabelian case. For the abelian case, the independent world-volume covariant couplings can be constructed using the pull-back metric. Subsequently, the coupling constants can be determined by imposing T-duality \cite{Karimi:2018vaf, Garousi:2022rcv}.
In the nonabelian case, it is known that the four YM field strength couplings coincide with the corresponding couplings in the nonabelian DBI action when a symmetric trace is imposed \cite{Tseytlin:1997csa}. However, it is also known that the couplings involving higher YM field strengths are not provided by the nonabelian symmetrized trace DBI action \cite{Hashimoto:1997gm}. Investigating the world-volume diffeomorphism symmetry for the nonabelian case would be intriguing, as it could help identify independent covariant and YM gauge invariant world-volume couplings. Subsequently, by imposing T-duality on these couplings, the coupling constants can be determined following a similar approach to what has been done in the context of the abelian case \cite{Karimi:2018vaf, Garousi:2022rcv}.

 
\vskip 1 cm
{\huge \bf Appendix: 5-derivative basis}
\vskip 0.5 cm
In this Appendix, we provide a step-by-step explanation of how, using the xAct package, we find the basis at the 5-derivative order in \reef{L5}.

We first consider all terms constructed from $\nabla\Phi$, $H$, $\cF$, $R$, and their derivatives, to construct the 5-derivative structures that have an odd number of $\cF$ and its derivatives. Here, $\Phi$ is the dilaton, $H$ has three totally antisymmetric indices, $R$ is the Riemann curvature, and $\cF$ has four indices - it is antisymmetric in the first two indices and also antisymmetric in the last two indices. All indices belong to a single manifold. 
We find that there are 62 such structures,
\beqa
L_0=\cF H^4+\cF^3H^2+\cF^5+\cF H^2R+\cdots\,.
\eeqa
We then add indices to these structures using the following commands:
\beqa
L_1&=&IndexFree[L_0];\nn\\
L_2&=&FromIndexFree[L_1]\,.
\eeqa 
Next, we consider all possible contractions of indices using the command:
\beqa
L_3=AllContractions[L_2]\,.
\eeqa
This produces many contractions in which the last two indices of $\cF$ contract with indices of $H$, $\nabla\Phi$, $R$, and their derivatives. These terms should be eliminated.

To do this, we introduce the YM tensor $F_{\alpha\beta}{}{}^{ij}$, which has two antisymmetric spacetime indices and two antisymmetric internal indices of the $SO(32)$ or $E_8 \times E_8$ spaces. We also introduce the tensor $E_\alpha{}^i$, which maps the spacetime indices to the internal space indices, with the property $E_\alpha{}^i E^\alpha{}^j = \kappa^{ij}$, where $\kappa^{ij}$ is the metric of the internal spaces.

By replacing $\cF_{\alpha\beta\mu\nu}$ in $L_3$ with $E_\mu{}^i E_\nu{}^j F_{\alpha\beta ij}$, using the identity $E_\alpha{}^i E^\alpha{}^j = \kappa^{ij}$, and removing all terms where the $E_\alpha{}^i$ remains, we find all contractions in $L_3$ where the last two indices of the $F$'s are contracted with each other. This results in 51 couplings, which we call $L_4$.
Finally, we use the command:
\beqa
L'=MakeAnsatz[L_4,ConstantPrefix, b']\,.
\eeqa
This adds coefficients $b_1', b_2', \cdots$ to the couplings, i.e.,
\beqa
L'=b'_{1}   F_{\alpha  }{}^{\epsilon  im } 
F^{\alpha  \beta  jk} F_{\beta  }{}^{\gamma  }{}_{i}{}^{l} F_{
\gamma  \delta  km } F_{\epsilon  }{}^{\delta  
}{}_{jl} +   \cdots\,.
\eeqa
There is also an overall dilaton factor of $e^{-2\Phi}$. The above represents all possible contractions between $\nabla\Phi$, $H$, $F$, $R$, and their derivatives at the 5-derivative order, which have an odd number of YM field strengths. However, they are not all independent. Some of them are interrelated through total derivative terms, while others are connected through field redefinitions or various Bianchi identities.

To eliminate the total derivative terms from the aforementioned couplings, the following terms are introduced into the Lagrangian $L'$:
\beqa
e^{-2\Phi} \mathcal{J}&\equiv&  \nabla_\alpha (e^{-2\Phi}{\cal I}^\alpha)\,.\labell{J3}
\eeqa
Here, the vector ${\cal I}^\alpha$ represents a collection of all possible covariant and gauge-invariant terms at the four-derivative level, including arbitrary parameters. The construction of such terms is similar to the construction of $L'$, except that ${\cal I}^\alpha$ has one spacetime index.

To eliminate the freedom of field redefinitions, it is necessary to perturb the metric, dilaton, $H$-field, and YM gauge fields in the two-derivative action which is given  in \reef{two2}. By utilizing the Bianchi identity satisfied by the $H$-field:
\beqa
\nabla_{[\alpha}H_{\beta\mu\nu]}+\frac{3}{4}F_{[\alpha\beta}{}^{ij}F_{\mu\nu]}{}_{ij}&=&0\,,\labell{iden0}
\eeqa
it is found that the perturbation of the $H$-field and the perturbation of the YM field are related through the equation $d(\delta H+3F^{ij}\delta A^{ij})=0$, where form notation is employed. Consequently, the perturbation of the $H$-field can be expressed as a linear combination of the YM gauge field perturbation and an arbitrary two-form $\delta B$:
\beqa
\delta H&=&3d\delta B-3F^{ij}\delta A_{ij}\,.
\eeqa 
Subsequently, the field redefinition introduces the following terms into the Lagrangian $L'$:
\beqa
\mathcal{K}
&\!\!\!\!\!\equiv\!\!\!\!\!\!& (\frac{1}{2} \nabla_{\gamma}H^{\alpha \beta \gamma} -  H^{\alpha \beta}{}_{\gamma} \nabla^{\gamma}\Phi)\delta B_{\alpha\beta}\nn\\
&&-(\nabla^{\beta}F_{\alpha\beta}{}^{ij}-2F_{\alpha\beta}{}^{ij}\nabla^{\beta}\Phi-\frac{1}{2}F^{\beta\mu ij} H_{\alpha\beta\mu})\delta A^{\alpha}{}_{ ij}\nn\\
&& -(  R^{\alpha \beta}-\frac{1}{4} H^{\alpha \gamma \delta} H^{\beta}{}_{\gamma \delta}+ 2 \nabla^{\beta}\nabla^{\alpha}\Phi-\frac{1}{2}F^{\alpha\mu ij} F^{\beta}{}_{\mu ij})\delta G_{\alpha\beta}\labell{eq.13}\\
\nn\\
&&-2( R\! -\!\frac{1}{12} H_{\alpha \beta \gamma} H^{\alpha \beta \gamma} \!+\! 4 \nabla_{\alpha}\nabla^{\alpha}\Phi \!-4 \nabla_{\alpha}\Phi \nabla^{\alpha}\Phi-\frac{1}{4}F_{\alpha\beta ij}F^{\alpha\beta ij}) (\delta\Phi-\frac{1}{4}\delta G^{\mu}{}_\mu) .\nn
\eeqa
In this expression, the perturbations $\delta G_{\mu\nu}, \delta B_{\mu\nu},\delta \Phi, \delta A_a{}^{ij}$ are constructed from the NS-NS and YM fields at the three-derivative order, with arbitrary coefficients.  $\delta G_{\mu\nu}$ is symmetric, and $\delta B_{\mu\nu}$ is antisymmetric.
The construction of these perturbations is similar to the construction of $L'$. By incorporating the total derivative terms and the contribution from field redefinitions into the Lagrangian $L'$, the resulting Lagrangian, denoted as $L$, exhibits different coupling constants $b_1,b_2,\cdots$. 
Consequently, the equation
\beqa
\Delta-{\cal J}-{\cal K}&=&0\,,\labell{DLK}
\eeqa
holds, where $\Delta = L -L'$ is equivalent to $L'$, but with coefficients $\delta b_1, \delta b_2, \ldots$, where $\delta b_i = b_i - b'_i$. 

To solve the equation \reef{DLK}, it is necessary to express it in terms of independent couplings by imposing the following Bianchi identities:
\beqa
 R_{\alpha[\beta\gamma\delta]}=0\,,&&
 \nabla_{[\mu}R_{\alpha\beta]\gamma\delta}=0\,,\,\,\,\,\,\,
{[}\nabla,\nabla{]}\mathcal{O}-R\mathcal{O}=0\,,\nn\\
\nabla_{[\alpha}H_{\beta\mu\nu]}+\frac{3}{4}F_{[\alpha\beta}{}^{ij}F_{\mu\nu]}{}_{ij}=0\,,&&
\nabla_{[\alpha}F_{\beta\gamma]}{}^{ij}=0\,.\labell{bian}
\eeqa
The above Bianchi identities can be imposed on the equation \reef{DLK} either in a gauge-invariant form or a non-gauge-invariant form. We impose them in a non-gauge-invariant form. Hence, we rewrite the terms in \reef{DLK} in a local frame where the covariant derivatives are expressed in terms of partial derivatives and the first partial derivative of the metric is zero. Additionally, the terms involving $H$ and $F$ and their derivatives in \reef{DLK} can be rewritten in terms of potentials using the relations \reef{FH}. The terms that have the YM gauge field without derivatives should be zero. This way, all the Bianchi identities are automatically satisfied.

The equation \reef{DLK} can then be solved to find relations between the $\delta b$'s, the parameters in the total derivative terms, and the parameters in the field redefinitions. We find 23 relations between only the $\delta b$'s that we are interested in, as well as some other relations that express the parameters in the total derivative terms and field redefinitions in terms of the $\delta b$'s. 
The 23 relations indicate that there are only 23 independent terms in $L'$. This means there are 28 terms in $L'$ that are removable by total derivative terms, field redefinitions, and Bianchi identities. 
If one removes such terms, then the basis becomes the minimal basis. There are different minimal schemes for choosing the 23 independent terms. To find these terms, one can choose the coefficients of 28 terms in $L'$ to be zero and then solve the equation \reef{DLK} again. If one finds 23 relations $\delta b_i = 0$, then that choice is allowed; otherwise, one should choose another set of 28 couplings to be zero. The couplings in \reef{L5} represent a particular minimal basis.
On the other hand, if one does not use field redefinitions, which means removing ${\cal K}$ from the equation \reef{DLK}, then one finds 31 relations between only the $\delta b$'s. This means the maximal basis has 31 independent terms. There are also different schemes for choosing the maximal basis.

Similar steps can be used to find the minimal and maximal bases at 7-derivative orders.

\end{document}